\documentclass[twocolumn]{aastex61}

\usepackage{color, CJKutf8}

\newcommand{\lnu}{\hat{L}_\nu}
\newcommand{\kanji}[1]{\begin{CJK}{UTF8}{mj}(#1)\end{CJK}}
\newcommand{\iso}[2]{{}^{#2}{\mbox{#1}}}

\received{---}
\revised{---}
\accepted{---}

\submitjournal{ApJL}

\shorttitle{The intermediate {\it r}-process in MR-SNe by the MRI}
\shortauthors{Nishimura et al.}

\begin{document}

\title{The intermediate \lowercase{r}-process in core-collapse supernovae driven by the magneto-rotational instability}

\correspondingauthor{Nobuya Nishimura}
\email{n.nishimura@keele.ac.uk, nishimura.nobuya@gmail.com}

\author[0000-0002-0842-7856]{N. Nishimura \kanji{西村信哉}}
\affil{Astrophysics Group, Faculty of Natural Sciences, Keele University, ST5 5BG Keele, UK}
\altaffiliation{UK Network for Bridging Disciplines of Galactic Chemical Evolution, www.bridgce.ac.uk, UK}

\author[0000-0002-4211-5784]{H. Sawai \kanji{澤井秀朋}}
\affiliation{Research Organization for Information Science \& Technology, 650-0047 Kobe, Japan}
\affiliation{Department of Physics, Waseda University, 169-8555 Tokyo, Japan}

\author[0000-0003-0304-9283]{T. Takiwaki \begin{CJK}{UTF8}{min}{(滝}\end{CJK}\hspace{-2.5pt}\begin{CJK}{UTF8}{mj}{脇知也)}\end{CJK}}
\affiliation{ Division of Theoretical Astronomy, National Astronomical Observatory of Japan,  181-8588 Mitaka, Japan}

\author{S. Yamada \kanji{山田章一}}
\affiliation{Department of Physics, Waseda University, 169-8555 Tokyo, Japan}

\author{F.-K. Thielemann}
\affiliation{Department of Physics, University of Basel, CH-4056 Basel, Switzerland}

\begin{abstract}
We investigated r-process nucleosynthesis in magneto-rotational supernovae, based on a new explosion mechanism induced by the magneto-rotational instability. A series of axisymmetric magneto-hydrodynamical simulations with detailed microphysics including neutrino heating is performed, numerically resolving the magneto-rotational instability. Neutrino-heating dominated explosions, enhanced by magnetic fields, showed mildly neutron-rich ejecta producing nuclei up to $A\sim130$ (i.e. the weak r-process), while explosion models with stronger magnetic fields reproduce a solar-like r-process pattern. More commonly seen abundance patterns in our models are in between the weak and regular r-process, producing lighter and intermediate mass nuclei. These {\it intermediate r-processes} exhibit a variety of abundance distributions, compatible with several abundance patterns in r-process-enhanced metal-poor stars. The amount of Eu ejecta $\sim 10^{-5} M_\odot$ in magnetically-driven jets agrees with predicted values in the chemical evolution of early galaxies. In contrast, neutrino-heating dominated explosions have a significant amount of Fe ($\iso{Ni}{56}$) and Zn, comparable to regular supernovae and hypernovae, respectively. These results indicate magneto-rotational supernovae can produce a wide range of heavy nuclei from iron-group to r-process elements, depending on the explosion dynamics.
\end{abstract}


\section{Introduction}
\setcounter{footnote}{5}

Core-collapse supernovae (CC-SNe) driven by rotation and magnetic fields, so-called {\it magneto-rotational supernovae} (MR-SNe), are a promising mechanism for several high-energy astronomical phenomena, e.g. magnetar formation, gravitational waves and hypernovae and gamma-ray bursts \citep[e.g.][]{2004ApJ...608..907Y, 2006PhRvD..74j4026S, 2009ApJ...691.1360T, 2014ApJ...785L..29M}. Jet-like explosions in CC-SNe are expected to eject very neutron-rich matter, appropriate for the r-process \citep[e.g.][]{2003ApJ...587..327C, 2012MNRAS.421.2763P}. In fact, recent studies \citep[][]{2012ApJ...750L..22W, 2015ApJ...810..109N}, based on multi-D magneto-hydrodynamics with sophisticated microphysics, confirmed that magnetically-driven jets produce heavy r-process elements.

MR-SNe may be rare compared with regular CC-SNe, as progenitors have rapid rotation, more frequently observed at low metallicities. The existence of fast rotating massive stars at early galaxies is also supported by detection of Ba and La in metal-poor stars \citep{2011Natur.472..454C}, which is explained by the enhanced s-process via strong rotational-induced mixing \citep{2012A&A...538L...2F, 2017arXiv170100489N}. Even if MR-SNe are only active in early galaxies, they can be responsible for the production of r-process elements by the entire CC-SNe in Galactic chemical evolution (GCE), because canonical CC-SNe produce only the lighter end of heavy nuclei in their proto-neutron star (proto-NS) winds \citep[e.g.][]{2013JPhG...40a3201A}.

Binary neutron star mergers (NSMs) are the most promising candidates of r-process sites \citep[e.g.][]{1999ApJ...525L.121F, 2011ApJ...738L..32G, 2012MNRAS.426.1940K, 2014ApJ...789L..39W}. However, there exist several unsolved problems for considering  NSMs as only r-process sources \citep[e.g.][]{2004A&A...416..997A}. \cite{2015MNRAS.452.1970W} and \cite{2015A&A...577A.139C} explained the chemical evolution of r-process nuclei, based on multiple sources, including NSMs and CC-SNe/MR-SNe. \cite{2015ApJ...811L..10T} showed that MR-SNe can explain the early growth of Eu in dwarf spheroidal galaxies for $[{\rm Fe}/{\rm H}] < -2$ by assuming an event rate of about $0.5\%$ of CC-SNe, while NSMs have problems to do so. Observation of ultra-faint galaxies \citep{2016AJ....151...82R, 2016Natur.531..610J} requires rare r-process events with large mass ejection, for which MR-SNe can be a source as well as NSMs \citep{2016ApJ...832..149B}.

\begin{figure}[t]
	\begin{center}
		\includegraphics[width=\hsize]{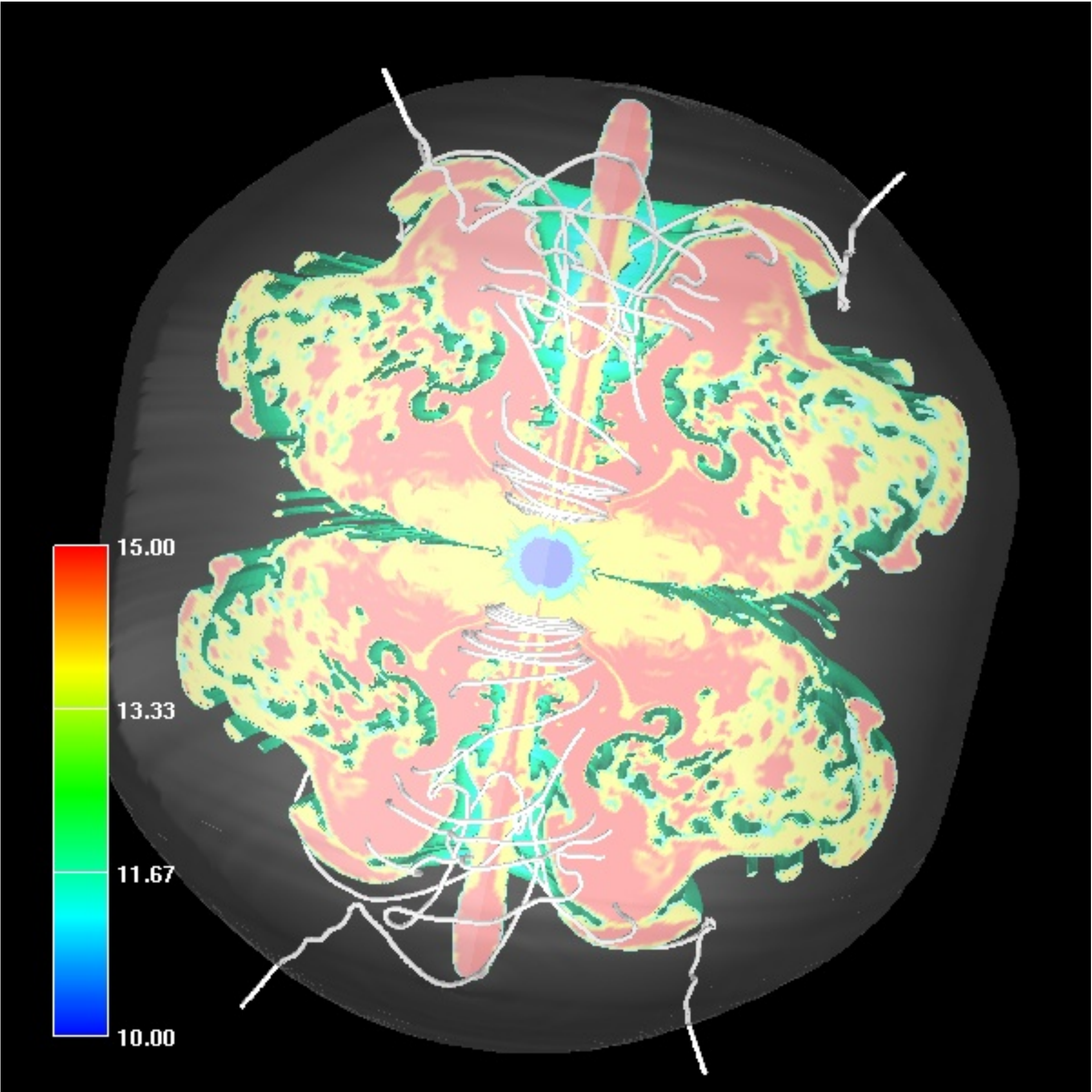}
		\caption{Entropy with magnetic field lines of an MR-SN model ($2000$~km range). The shock from is illustrated by the surrounded white surface. The color of entropy is apparently different from the color scale ($10$--$15~k_{\rm B}~{\rm baryon}^{-1}$) in visualization.}
	\label{fig-mrsn}
	\end{center}
\end{figure}

A remaining problem of MR-SNe is mechanisms of magnetic-field enhancement during collapse. The most promising process is the magneto-rotational instability (MRI), which converts rotation energy into magnetic energy. The MRI in proto-NS cores has been investigated on several scales (with related limitations) from local boxes \citep{2009A&A...498..241O, 2012ApJ...759..110M, 2016MNRAS.456.3782R} to global scales \citep{2013ApJ...770L..19S, 2015Natur.528..376M}. \cite{2014ApJ...784L..10S, 2016ApJ...817..153S}, based on long-term global MHD simulations in axisymmetry, found a new explosion mechanism influenced by the MRI. Besides magnetically driven polar jets, the explosion takes place in all directions, for which a typical dynamical structure is shown in Figure~\ref{fig-mrsn} (see, Section~\ref{sec-mrsn} for details).

In this Letter, we present the results of r-process nucleosynthesis in MR-SNe, based on the MRI-driven explosion mechanism. In Section~\ref{sec-mrsn}, we perform a series of simulations of MR-SNe resolving the MRI, mostly focusing on the outer layers of the proto-NS. We consider the effect of neutrino heating in explosion dynamics, extended from \cite{2014ApJ...784L..10S, 2016ApJ...817..153S}. In Section~\ref{sec-rproc}, results of nucleosynthesis for all explosion models are shown with comparison to observed r-process abundances.

\begin{figure*}[t]
	\begin{center}
		\includegraphics[width=0.9\hsize]{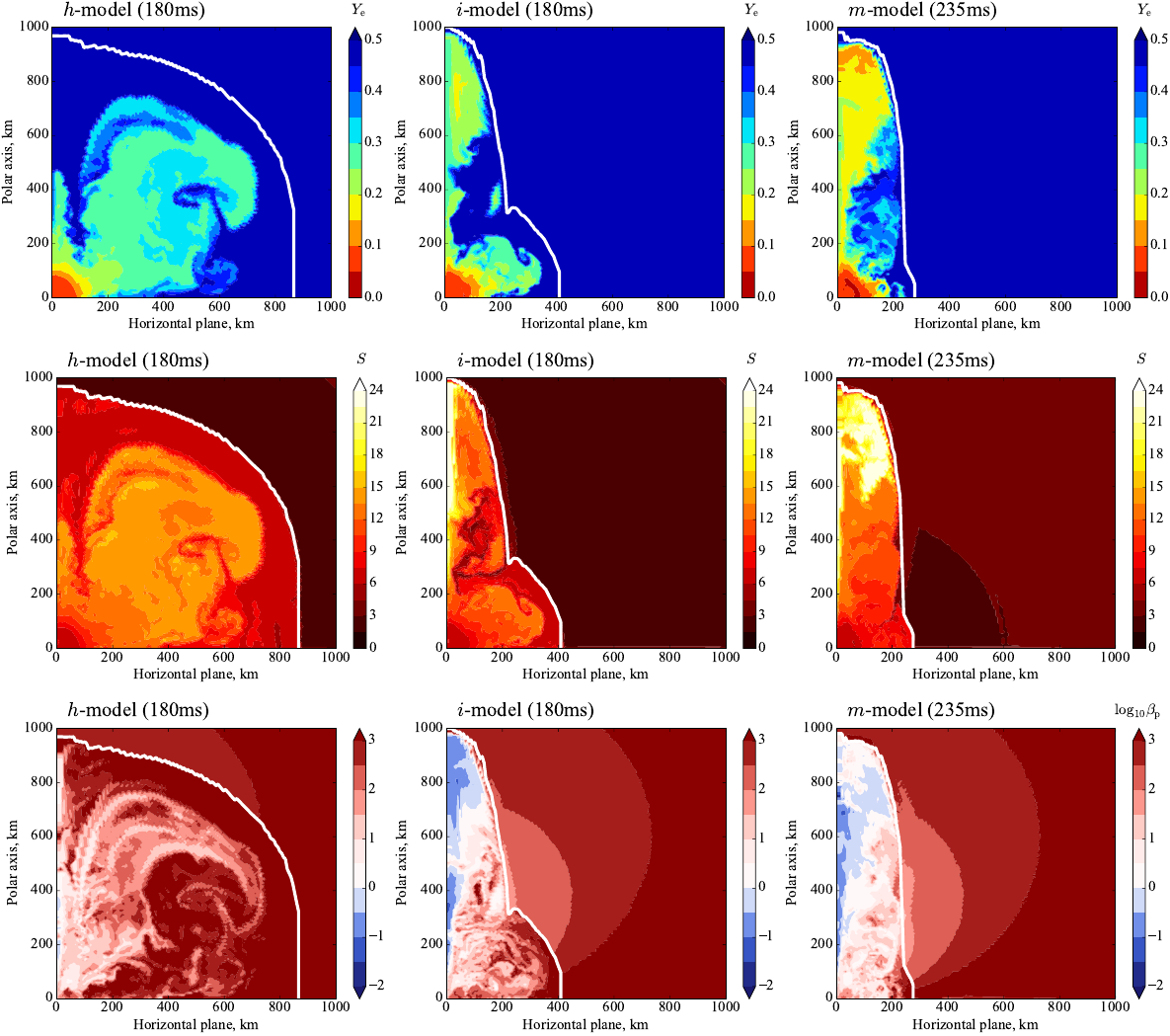}
		\caption{Dynamics of the explosion models when the shock front (the white line) reaches $\sim 1000$~km in ms after the bounce. Distributions of plasma $\beta$ ($\beta_{\rm p}$), entropy and $Y_{\rm e}$ (from top to bottom) are plotted for $h$-, $i$- and $m$-models.}
	\label{fig-hydro}
	\end{center}
\end{figure*}

\section{MRI-driven Core-collapse Supernovae\label{sec-mrsn}}

We perform hydrodynamical simulations for MR-SNe with an MHD code, YAMAZAKURA\footnote{The name derives from ``wild cherry blossoms'' in Japanese.} \citep[][]{2013ApJ...770L..19S}. As we are based on a $15M_\odot$ progenitor model \citep[][]{1995ApJS..101..181W} calculated in 1D (spherical symmetry), we consider initial rotation and magnetic fields using analytic formulae. We adopt shellular rotation by $\Omega(r) = \Omega_{0} {r_0}^2/({r_0}^2 + r^2)$, where $r$ is the distance from the center. We choose $r_0 = 1,000$~km (approximately the size of the iron core) and $\Omega_{0}=2.7$ rad~s$^{-1}$, leading to millisecond rotation after the collapse. For the initial magnetic fields, we apply the same dipole-like configuration as \cite{2016ApJ...817..153S} with a maximum value of $2 \times 10^{11}$~G around the center, while the value decreases to $1 \times 10^{11}$~G at the edge of the core ($\sim 1000$~km). At the surface of $1.4 M_\odot$ in the enclosed mass, the magnetic flux is $7.0 \times 10^{27} {\rm cm}^2~{\rm G}$, comparable to those of magnetar candidates.

We include neutrino heating in dynamics by the light-bulb method, treating the proto-NS as a point source, with a simplified neutrino-emission model. Using the same initial conditions, we calculated the time-evolution of neutrinos calculated by another supernova code \citep{2016MNRAS.461L.112T} with an advanced neutrino transport scheme based on IDSA, where we denote $L_{\nu}^{\rm IDSA}$ for the neutrino luminosity. We assumed $4$ and $6$ MeV for the temperatures of electron and anti-electron neutrinos, respectively, corresponding to the average values of the IDSA simulation.

\begin{figure*}[]
	\begin{center}
		\includegraphics[width=0.9\hsize]{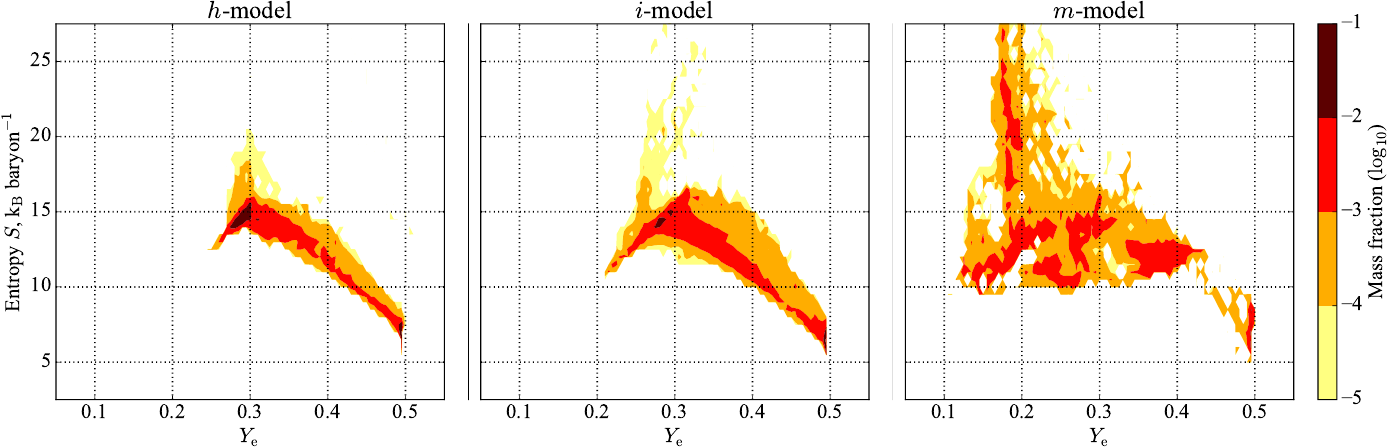}
\caption{The mass fraction (in logarithmic scale) of ejecta on the $Y_{\rm e}$--$S$ (entropy) plane.}
	\label{fig-yes}
	\end{center}
\end{figure*}

As in \cite{2014ApJ...784L..10S}, we initially perform a low-resolution simulation, ignoring the MRI, in a large spatial region, i.e. a $4000$~km radius area covered by $720 (r) \times 60 (\theta)$ grids in spherical coordinates. A higher resolution calculation, resolving the MRI, is conducted from $1$~ms after the bounce. To reduce the computational time, we limit the high-resolution region to the radial range of $30$--$2000$~km with the boundaries obtained by the low-resolution simulation. With this simplification, we resolve the MRI in the outer layers of the proto-NS \citep[see,][for more details]{2016ApJ...817..153S}, which is the most important for of the explosion. High-resolution simulations, resolving the MRI, have a $2100 (r) \times 800 (\theta)$ mesh with the $60$~m innermost grid size.

The structure of the explosion model, illustrating entropy with magnetic field lines (in 3D), is shown in Figure~\ref{fig-mrsn}. A bipolar jet along the rotational axis is launched due to magnetic pressure with entropies beyond $15~k_{\rm B}~{\rm baryon}^{-1}$. This jet-like explosion is wrapped by magnetic field lines, as commonly seen in previous magnetically-driven MR-SN models. We also see chaotic convective motion in the off-rotation-axis region with complicated entropy distribution.

The explosion process of the polar jet is similar to the magnetically dominated mechanism in previous studies \citep[e.g.][]{2012ApJ...750L..22W, 2015ApJ...810..109N}, where strong magnetic pressure drives outward ejection, overcoming ram pressure. In contrast, the explosion in other directions is driven by neutrino heating, where the MRI significantly enhances the convection and angular momentum transport \citep{2014ApJ...784L..10S}. Although neutrino heating plays an important role, neutron-rich matter is ejected especially in polar directions with a smaller influence of $\nu_e$-capture on neutrons due to shorter expansion/explosion timescales.

\begin{figure}[t]
	\begin{center}
		\includegraphics[width=1.0\hsize]{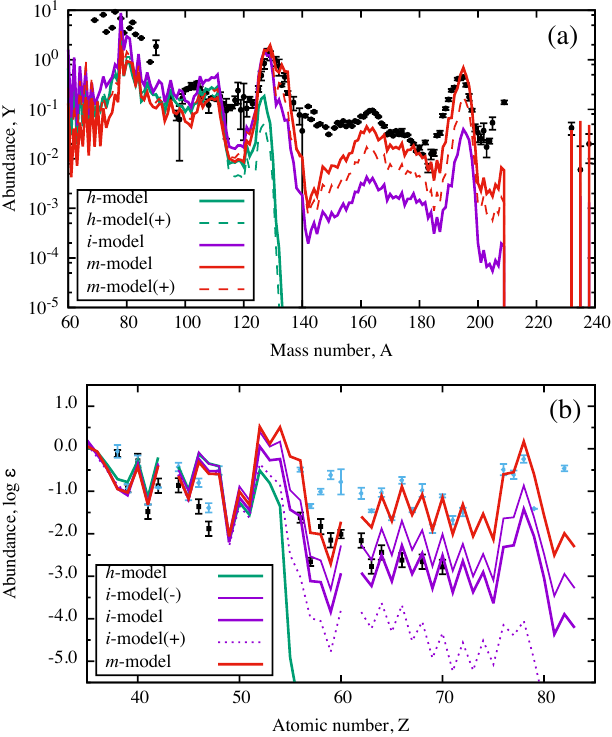}
		\caption{The final abundances of nucleosynthesis calculations: (a) compared with the solar abundances \citep{1999ApJ...525..886A}; (b) metal-poor stars, i.e. HD122563 \citep[][]{2006ApJ...643.1180H} and  CS22892-052 \citep[][]{1996ApJ...467..819S} denoted by black and cyan dots, respectively, where abundances are normalized for $Z=40$ of HD122563.}
	\label{fig-rproc}
	\end{center}
\end{figure}

We compute additional models to investigate the effect of magnetic fields in explosion dynamics. However, simulating MRI-driven explosion models with different initial magnetic fields requires a huge amount of computer resources, because the saturation of the MRI is strongly depending on the magnetic fields, i.e. all explosion models have different criteria for numerical convergence \citep[see Figure 6 of ][]{2016ApJ...817..153S}. In the current study, therefore, we adopt a more simplified parametric method based on the above MRI-driven explosion model (shown in Figure~\ref{fig-mrsn}), whose numerical convergence has been confirmed.\footnote{We confirmed that a higher resolution model, of which the finest grid is $30$~m, shows the same result of r-process nucleosynthesis (plotted in Figure~\ref{fig-rproc}) as well as explosion dynamics.} Instead of changing magnetic fields, we vary the time evolution of neutrino luminosity by multiplying a scale factor as $\lnu \equiv L_{\nu}/{L_{\nu}}^{\rm IDSA} = 0.10, 0.20, 0.30, 0.40, 0.50, 0.60, 0.75$ and $1.25$.

We suppose that a lower $\lnu$ model has a stronger influence of magnetic fields. To evaluate the strength of magnetic fields in jets, we consider the minimum value of the ratio of gas to magnetic pressure ($\beta_{\rm p}$), i.e. $\beta_{\rm p,min}$. Since the $\beta_{\rm p,min}$ has a large fluctuation, we calculate the {\it mean value} denoted by $\langle \beta_{\rm p, min} \rangle$ during jet propagation (from jet launching to the time the shock front reaches $2,000$~km). A lower $\langle \beta_{\rm p, min} \rangle$ is obtained in models with lower $\lnu$, where $\langle \beta_{\rm p, min} \rangle = 0.027, 0.038, 0.047, 0.060, 0.071, 0.083, 0.15, 0.20$ and $0.30$ in ascending order of $\lnu$. We expect neutron-rich material is mostly ejected in magnetically-driven jets rather than heating-driven ejecta. In the jet direction, neutrino emission can be weakened due to lower mass accretion onto the proto-NS. Adopting a lower $\lnu$, therefore, is consistent with decreasing $\lnu$ in the polar direction.

We label the standard MRI-driven explosion model with $\lnu = 1.0$ as the $h$-model, standing for neutrino {\it heating} to be dominant, while the case of {\it magnetically}-driven jet with $\lnu = 0.20$ is named as the $m$-model. We choose the case of $\lnu = 0.60$ as a typical {\it intermediate} explosion model as the $i$-model. As the behaviour of explosion varies gradually depending on $\lnu$, the distinction of models has uncertainties, e.g. the $m$-model can be categorized in the $i$-model as an extreme case. We adopt $\lnu = 0.40$ and $0.75$ models as a variation of the $i$-model, denoted by the $i$-model(-) and $i$-model(+), respectively. The $h$-model(+) and $m$-model(+), based on $\lnu = 1.25$ and $0.40$, respectively, are also referred to discuss the impact of neutrino absorption.
 
The $Y_{\rm e}$, entropy ($S$) and $\beta_{\rm p}$ for the selected models are shown in Figure~\ref{fig-hydro}, when the shock front reaches $\sim 1000$~km. For the $m$-model, we clearly see a magnetically-driven dipole jet with lower $\beta_{\rm p}$ (i.e. high magnetic pressure), which shows very neutron-rich ejecta and $Y_{\rm e} \sim 0.2$. The $h$-model has a weaker jet, which is almost negligible, and the dominant component consists of the convective motion driven by neutrino heating. The physical properties of the $i$-model are intermediate between both the above models, a combination of magnetically-jets and neutrino-heating caused explosions. The $i$-model shows a relatively higher $Y_{\rm e}$, but the ejecta is still neutron-rich ($Y_{\rm e} < 0.4$).

We calculate the time evolution of ejected matter via Lagrangian tracer particles \citep[see, Section~2.3 of][]{2015ApJ...810..109N}. Figure~\ref{fig-yes} shows the mass fraction of ejecta in the $Y_{\rm e}$--$S$ plane. Since the entropy is relatively low, the resulting r-process is strongly dependent on $Y_{\rm e}$. All explosion models including the $h$-model have neutron-rich ejecta ($Y_{\rm e} < 0.4$), missing in regular CC-SNe.

\section{Nucleosynthesis}
\label{sec-rproc}

We perform nucleosynthesis calculation using a nuclear reaction network code \citep{2015ApJ...810..109N, 2016PhLB..756..273N}, with reaction rate data taken mostly from \cite{2000ADNDT..75....1R}. Theoretical reaction rates (e.g. for neutron-capture) for neutron-rich nuclei are based on mass predictions by \cite{1995ADNDT..59..185M} and the theoretical $\beta$-decay rates are taken from \cite{2003PhRvC..67e5802M}. The impact of $\beta$-decay was discussed in \cite{2012PhRvC..85d8801N, 2016PhLB..756..273N}.

Results of nucleosynthesis calculations are shown in Figure~\ref{fig-rproc}a for selected models, comparing with the solar abundances \citep[s-process residuals by][]{1999ApJ...525..886A}. The calculated abundance patterns vary in the $A > 130$ region according to the predicted $Y_{\rm e}$--$S$ distributions (Figure~\ref{fig-yes}). The r-process in low entropy conditions can be sensitive to self nuclear heating, which is not considered in the present calculations. Although final abundances may be modified as shown by \cite{2016MNRAS.463.2323W}, based on similar physical ($Y_{\rm e}$--$S$) environments, the variation due to different explosion dynamics (i.e. $h$-, $i$- and $m$-models) is more significant.

The $m$-model with sufficiently high enough magnetic fields reproduces the solar r-process abundances, as shown in \cite{2012ApJ...750L..22W} and ``prompt-jets'' of \cite{2015ApJ...810..109N}, while the $m$-model(+) with a higher $\lnu$ shows underproduction for nuclei heavier than the second peak ($A \sim 130$). We see significant deficiencies in calculated abundances around the second peak, as appeared in previous studies \citep{2012ApJ...750L..22W, 2015ApJ...810..109N} for MR-SN models. This may be caused by the defect of theoretical nuclear reaction/decay rates (e.g. $\beta$-decay rates and fission fragments) rather than astrophysical models (hydrodynamical environments). In fact, \cite{2014ApJ...792....6K} showed that updated $\beta$-decay rates, based on the latest FRDM, improve the production of rare-earth nuclei. It has also been shown that other theoretical $\beta$-decay rates provide different abundance features \citep[e.g.][]{2011ApJ...738L..32G, 2014ApJ...789L..39W, 2016MNRAS.463.2323W}. In addition, \cite{2015ApJ...808...30E} showed that their new fission fragments possibly improve the production of nuclei in the rare-earth region.

The $h$-model shows production up to the second peak by a ``failed r-process'' with insufficient neutron-rich ejecta \citep[similar yields of ``delayed-jets'' in][but the ejection process is different]{2015ApJ...810..109N}. The $h$-model(+) shows slightly less $A \sim 130$ peak production, suppressed by stronger neutrino absorption. Nevertheless, a significant amount of lighter r-process nuclei is ejected in the $h$-model, which is difficult in the proto-NS winds of CC-SNe \citep{2013JPhG...40a3201A}.

The $i$-model exhibits different nucleosynthesis features, which the abundances of heavy nuclei are between the $m$- and $h$-models, as expected by the $Y_{\rm e}$--$S$ distribution. Another comparison with r-process enhanced metal-poor stars is shown in Figure~\ref{fig-rproc}b. The results of the $i$-model(-) and $i$-model(+) are added as variation of the $i$-model. These models show abundance patterns close to weak r abundances \citep{2006ApJ...643.1180H} rather than a solar-like pattern \citep{1996ApJ...467..819S}. In the region of $Z>60$ elements, we can see that the abundances increase monotonically as the $\lnu$ decreases.

\begin{figure}[t]
	\begin{center}
		\includegraphics[width=1.0\hsize]{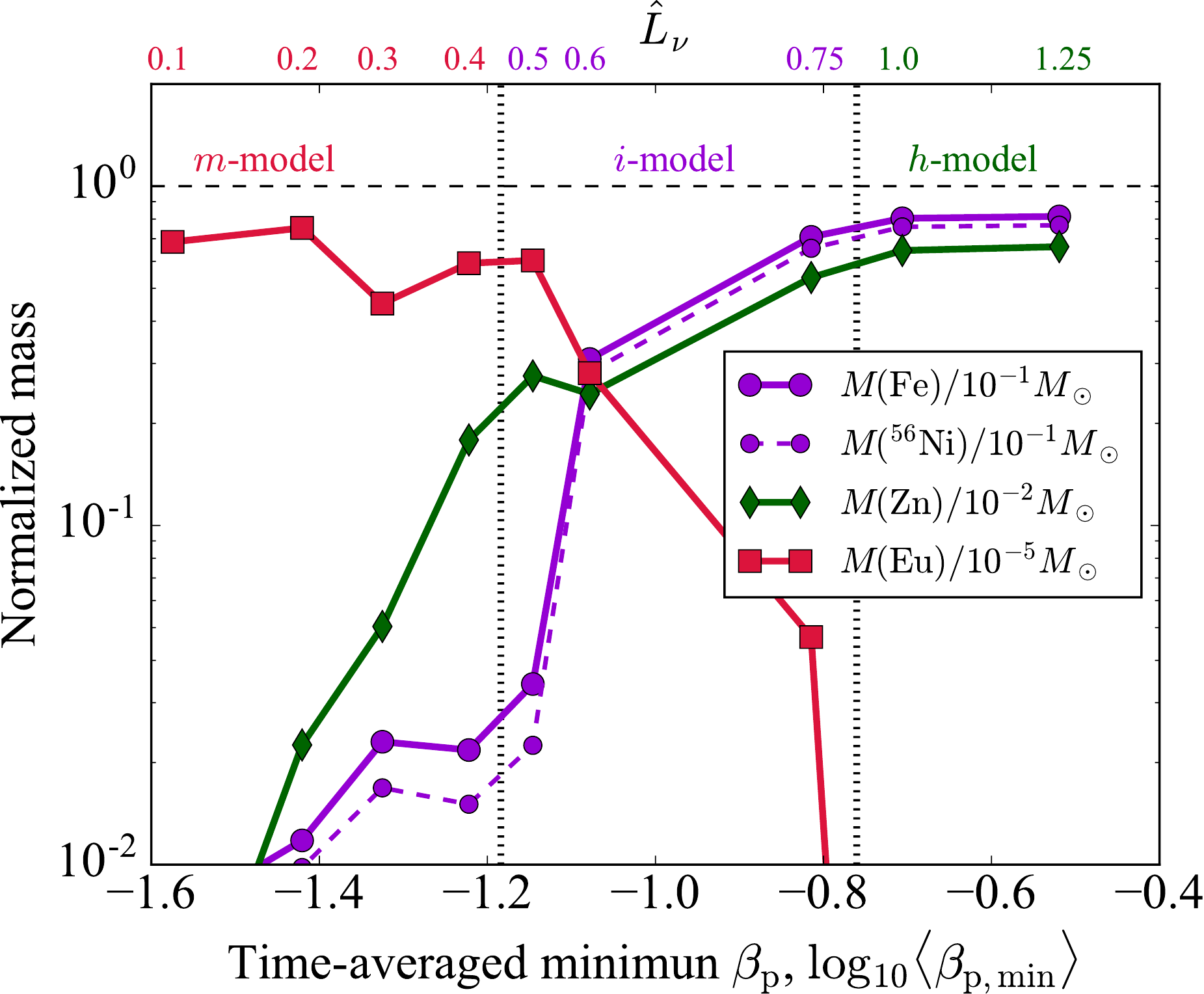}
		\caption{Ejected masses of Fe, $^{56}{\rm Ni}$ (before decay), Zn and Eu, normalized by 0.1, 0.1, $10^{-2}$ and $10^{-5} M_\odot$, respectively, as a function of $\langle \beta_{\rm p, min}\rangle$ with corresponding $\lnu$ (top).}
	\label{fig-nieu}
	\end{center}
\end{figure}

In Figure~\ref{fig-nieu}, the ejected masses of Fe, $^{56}{\rm Ni}$ (before $\beta$ decay) and Zn (representing trans-iron elements) are plotted together with the mass of Eu ejecta as a function of $\langle \beta_{\rm p,  min} \rangle$ as well as $\lnu$. The masses of Fe, $^{56}{\rm Ni}$ and Zn are normalised by $0.1$, $0.1$ and $10^{-2} M_\odot$, respectively, which are typical scales of CC-SNe, and Eu is normalized by $10^{-5}M_\odot$, suggested by \cite{2015ApJ...811L..10T}. The $h$-model has $0.076M_\odot$ $^{56}{\rm Ni}$ ejecta, which is coincidentally a similar value estimated for SN1987A. This indicates that MR-SNe (for a certain parameter range) are optically observed as canonical CC-SNe as well as the source of lighter r-process nuclei. Strong magnetically-driven models ($\lnu <0.5$) including the $m$-model shows higher amount of Eu ($\sim 0.1M_\odot$) with less dependence on explosion parameters.

Even explosion models with a larger $^{56}$Ni mass have lighter r-process yields, and the value of [Eu/Fe] varies from $-2.85$ ($h$-model), $2.45$ ($i$-model) to $4.30$ ($m$-model), consistent with dispersion in early galaxies. Additionally, the production of Zn is significant for all models, where ${\rm [Zn/Fe]} = 1.69$, $1.67$ and $2.05$ for $h$-, $i$- and $m$-models, respectively. Although the amount of Zn varies depending on $\lnu$, all models have a larger value than regular CC-SNe, for which previous explosion models based on simplified central engines (neglecting neutrino interactions) underproduced Zn \citep[see,][]{2006PhRvL..96n2502F}. These values are comparable to those for HNe \citep[see, a resent review,][]{2013ARA&A..51..457N}.

The anti-correlation between Fe ($\iso{Ni}{56}$) and Eu is physically obvious by the $Y_{\rm e}$-distribution of ejecta (Figure~\ref{fig-yes}). Iron-group elements are produced in a higher $Y_{\rm e} \sim 0.5$, while r-process elements are produced in a lower $Y_{\rm e} < 0.3$ ejecta. The production of Zn is correlated with Fe for $h$- and $i$-models, in contrast, the production of $\iso{Zn}{66, 68}$ is significant for lower $\lnu$ values, including the $m$-model, because they experience mildly neutron-rich ejecta with $Y_{\rm e} = 0.4$--$0.45$.

Although the $Y_{\rm e}$ of ejecta has an uncertainty due to the simplified neutrino transport method, our conclusion may not be significantly modified. Fe/$\iso{Ni}{56}$ and Eu (and r-process elements) are produced in environments with less neutrino absorption, i.e. $\iso{Ni}{56}$ is produced in the outer layer; Eu is produced in the jet with fast expansion velocity \citep{2015ApJ...810..109N}. On the other hand, Zn is synthesized in $Y_{\rm e} = 0.4$--$0.45$ ejecta in the inner layer, which is more sensitive to neutrino absorption. However, our explosion models including the $m$-model have wider $Y_{\rm e}$ distribution including $Y_{\rm e} < 0.4$ ejecta (Figure~\ref{fig-yes}), so that a certain amount of Zn-rich ejecta may remain under the significant effect of neutrino absorption, where the $Y_{\rm e}$ distribution shifts toward a larger value.

\section{Summary and Discussion}
\label{sec-conclusion}

We investigated r-process nucleosynthesis in MR-SNe, driven by the MRI. We found that an r-process, producing $A > 100$ nuclei with various abundance patters, takes place, depending on the strength of magnetic pressure (denoted by $\langle \beta_{\rm p, min}\rangle$) in the magnetically-driven jet. Explosion models, in which neutrino heating and magnetic fields are comparable (the $i$-models), produce a varying range of abundance patterns between the weak r-process and the solar-abundances.

Our finding is that these $i$-models can explain a weak r-process pattern and lead to a variety of abundance patterns among heavier nuclei. This behaviour could explain the diversity of r-process abundances observed in metal-poor stars, if originating from MR-SNe contributions. The conditions range from the extreme cases of $m$-models, which produce the full r-process abundances up to the heaviest nuclei, down to the $i$- or $h$-models, responsible for weak r-process patterns as seen in ``Honda-type'' r-process patterns \citep[][]{2004ApJ...607..474H} or only nuclei up to the second peak ($A=130$). More detailed observations are desirable to see if the low-metallicity behaviour reflects ejecta compositions of such individual events or whether we see a superposition of extreme cases of strong and weak r-process nucleosynthesis environments. Although we currently have a few examples for irregular r-process patterns, recent investigation suggest more of such cases \citep[][and private communication]{2014AIPC.1594..123A}.

We calculated ejected masses for iron-group and trans-iron nuclei as well as r-process nuclei. While magnetically-driven jets show a lower $\iso{Ni}{56}$ mass, explosions with stronger neutrino-heating eject amounts of $\iso{Ni}{56}$ comparable to regular CC-SNe. All our models show significant production of ${\rm [Zn/Fe]} > 1.5$, however, in this case, the dominant isotopes are neutron-rich $\iso{Zn}{66, 68}$ (rather than $\iso{Zn}{64}$), and [Eu/Fe] shows a large dispersion. These nucleosynthetic properties are clearly different from canonical CC-SNe and can be compatible to HNe.

MR-SNe are basically 3D phenomena, of which the launched polar-jet as in \cite{2012ApJ...750L..22W}, can be destroyed by hydrodynamical instabilities. \cite{2014ApJ...785L..29M} showed that the kink-instability deforms the ejection of a magnetically jet and significantly change dynamics in early phase of explosion, although neutron-rich matter ($Y_{\rm e} = 0.1$--$0.2$) is still expected. The properties of ejected neutron-rich matter shown in the current study possibly change in 3D simulations. Thus, further studies for the effect of the MRI in full global 3D simulations are important, which current simulations \citep[e.g.][]{2015Natur.528..376M} are limited to early phases (the inner region). On the other hand, as 3D simulations require a huge amount of computational resources, more systematic studies with a wide range of parameters for rotation and magnetic fields are desirable, even within the axis-symmetric MHD framework. The transition of r-process abundances from weak to strong r-process patterns should be understood as a function of stellar rotation and magnetic fields.

Besides the large scatter of [Eu/Fe] in abundance observations of low-metallicity stars, witnessing GCE, there exist additional indications that the astrophysical r-process is a rare event in comparison to regular CC-SNe. The Pu-content in deep-sea sediments \citep{2015NatCo...6E5956W} can be explained by NS-mergers \citep{2015NatPh..11.1042H}, but would also be consistent with MR-SNe. \cite{2016ApJ...827...83K} showed problems to explain r-process nuclei in Galactic cosmic rays with present astrophysical scenarios. We expect that progress in r-process predictions for MR-SNe as well as NS mergers will shed light on these open questions. 

The numerical data of nucleosynthesis yields and trajectories are available at \url{http://github.com/nnobuya/mrsn}.

\acknowledgments
The authors thank the referee for his/her valuable comments for improving the manuscript. This project was supported by the ERC (EU-FP7-ERC-2012-St Grant 306901 SHYNE, EU-FP7-ERC Advanced Grant 321263 FISH), JSPS (16H03986, 24103006, 24244036, 26800149, 26870823) and MEXT (15H01039, 15H00789). T.T. and S.Y. were supported by MEXT as ``Priority Issue on Post-K computer'' (Elucidation of the Fundamental Laws and Evolution of the Universe) and JICFuS. Parts of the computations were carried out on XC30 and PC cluster at CfCA, National Astronomical Observatory of Japan.



\begin{thebibliography}{}
\expandafter\ifx\csname natexlab\endcsname\relax\def\natexlab#1{#1}\fi
\providecommand{\url}[1]{\href{#1}{#1}}

\bibitem[{{Aoki} {et~al.}(2014){Aoki}, {Aoki}, {Ishimaru}, \&
  {Wanajo}}]{2014AIPC.1594..123A}
{Aoki}, M., {Aoki}, W., {Ishimaru}, Y., \& {Wanajo}, S. 2014, in American
  Institute of Physics Conference Series, Vol. 1594, American Institute of
  Physics Conference Series, ed. S.~{Jeong}, N.~{Imai}, H.~{Miyatake}, \&
  T.~{Kajino}, 123--128

\bibitem[{{Arcones} \& {Thielemann}(2013)}]{2013JPhG...40a3201A}
{Arcones}, A., \& {Thielemann}, F.-K. 2013, Journal of Physics G Nuclear
  Physics, 40, 013201

\bibitem[{{Argast} {et~al.}(2004){Argast}, {Samland}, {Thielemann}, \&
  {Qian}}]{2004A&A...416..997A}
{Argast}, D., {Samland}, M., {Thielemann}, F.-K., \& {Qian}, Y.-Z. 2004, \aap,
  416, 997

\bibitem[{{Arlandini} {et~al.}(1999){Arlandini}, {K{\"a}ppeler}, {Wisshak},
  {Gallino}, {Lugaro}, {Busso}, \& {Straniero}}]{1999ApJ...525..886A}
{Arlandini}, C., {K{\"a}ppeler}, F., {Wisshak}, K., {et~al.} 1999, \apj, 525,
  886

\bibitem[{{Beniamini} {et~al.}(2016){Beniamini}, {Hotokezaka}, \&
  {Piran}}]{2016ApJ...832..149B}
{Beniamini}, P., {Hotokezaka}, K., \& {Piran}, T. 2016, \apj, 832, 149

\bibitem[{{Cameron}(2003)}]{2003ApJ...587..327C}
{Cameron}, A.~G.~W. 2003, \apj, 587, 327

\bibitem[{{Cescutti} {et~al.}(2015){Cescutti}, {Romano}, {Matteucci},
  {Chiappini}, \& {Hirschi}}]{2015A&A...577A.139C}
{Cescutti}, G., {Romano}, D., {Matteucci}, F., {Chiappini}, C., \& {Hirschi},
  R. 2015, \aap, 577, A139

\bibitem[{{Chiappini} {et~al.}(2011){Chiappini}, {Frischknecht}, {Meynet},
  {Hirschi}, {Barbuy}, {Pignatari}, {Decressin}, \&
  {Maeder}}]{2011Natur.472..454C}
{Chiappini}, C., {Frischknecht}, U., {Meynet}, G., {et~al.} 2011, \nat, 472,
  454

\bibitem[{{Eichler} {et~al.}(2015){Eichler}, {Arcones}, {Kelic}, {Korobkin},
  {Langanke}, {Marketin}, {Martinez-Pinedo}, {Panov}, {Rauscher}, {Rosswog},
  {Winteler}, {Zinner}, \& {Thielemann}}]{2015ApJ...808...30E}
{Eichler}, M., {Arcones}, A., {Kelic}, A., {et~al.} 2015, \apj, 808, 30

\bibitem[{{Freiburghaus} {et~al.}(1999){Freiburghaus}, {Rosswog}, \&
  {Thielemann}}]{1999ApJ...525L.121F}
{Freiburghaus}, C., {Rosswog}, S., \& {Thielemann}, F.-K. 1999, \apjl, 525,
  L121

\bibitem[{{Frischknecht} {et~al.}(2012){Frischknecht}, {Hirschi}, \&
  {Thielemann}}]{2012A&A...538L...2F}
{Frischknecht}, U., {Hirschi}, R., \& {Thielemann}, F.-K. 2012, \aap, 538, L2

\bibitem[{{Fr{\"o}hlich} {et~al.}(2006){Fr{\"o}hlich},
  {Mart{\'{\i}}nez-Pinedo}, {Liebend{\"o}rfer}, {Thielemann}, {Bravo}, {Hix},
  {Langanke}, \& {Zinner}}]{2006PhRvL..96n2502F}
{Fr{\"o}hlich}, C., {Mart{\'{\i}}nez-Pinedo}, G., {Liebend{\"o}rfer}, M.,
  {et~al.} 2006, Physical Review Letters, 96, 142502

\bibitem[{{Goriely} {et~al.}(2011){Goriely}, {Bauswein}, \&
  {Janka}}]{2011ApJ...738L..32G}
{Goriely}, S., {Bauswein}, A., \& {Janka}, H.-T. 2011, \apjl, 738, L32

\bibitem[{{Honda} {et~al.}(2006){Honda}, {Aoki}, {Ishimaru}, {Wanajo}, \&
  {Ryan}}]{2006ApJ...643.1180H}
{Honda}, S., {Aoki}, W., {Ishimaru}, Y., {Wanajo}, S., \& {Ryan}, S.~G. 2006,
  \apj, 643, 1180

\bibitem[{{Honda} {et~al.}(2004){Honda}, {Aoki}, {Kajino}, {Ando}, {Beers},
  {Izumiura}, {Sadakane}, \& {Takada-Hidai}}]{2004ApJ...607..474H}
{Honda}, S., {Aoki}, W., {Kajino}, T., {et~al.} 2004, \apj, 607, 474

\bibitem[{{Hotokezaka} {et~al.}(2015){Hotokezaka}, {Piran}, \&
  {Paul}}]{2015NatPh..11.1042H}
{Hotokezaka}, K., {Piran}, T., \& {Paul}, M. 2015, Nature Physics, 11, 1042

\bibitem[{{Ji} {et~al.}(2016){Ji}, {Frebel}, {Chiti}, \&
  {Simon}}]{2016Natur.531..610J}
{Ji}, A.~P., {Frebel}, A., {Chiti}, A., \& {Simon}, J.~D. 2016, \nat, 531, 610

\bibitem[{{Korobkin} {et~al.}(2012){Korobkin}, {Rosswog}, {Arcones}, \&
  {Winteler}}]{2012MNRAS.426.1940K}
{Korobkin}, O., {Rosswog}, S., {Arcones}, A., \& {Winteler}, C. 2012, \mnras,
  426, 1940

\bibitem[{{Kratz} {et~al.}(2014){Kratz}, {Farouqi}, \&
  {M{\"o}ller}}]{2014ApJ...792....6K}
{Kratz}, K.-L., {Farouqi}, K., \& {M{\"o}ller}, P. 2014, \apj, 792, 6

\bibitem[{{Kyutoku} \& {Ioka}(2016)}]{2016ApJ...827...83K}
{Kyutoku}, K., \& {Ioka}, K. 2016, \apj, 827, 83

\bibitem[{{Masada} {et~al.}(2012){Masada}, {Takiwaki}, {Kotake}, \&
  {Sano}}]{2012ApJ...759..110M}
{Masada}, Y., {Takiwaki}, T., {Kotake}, K., \& {Sano}, T. 2012, \apj, 759, 110

\bibitem[{{M{\"o}ller} {et~al.}(1995){M{\"o}ller}, {Nix}, {Myers}, \&
  {Swiatecki}}]{1995ADNDT..59..185M}
{M{\"o}ller}, P., {Nix}, J.~R., {Myers}, W.~D., \& {Swiatecki}, W.~J. 1995,
  Atomic Data and Nuclear Data Tables, 59, 185

\bibitem[{{M{\"o}ller} {et~al.}(2003){M{\"o}ller}, {Pfeiffer}, \&
  {Kratz}}]{2003PhRvC..67e5802M}
{M{\"o}ller}, P., {Pfeiffer}, B., \& {Kratz}, K.-L. 2003, \prc, 67, 055802

\bibitem[{{M{\"o}sta} {et~al.}(2015){M{\"o}sta}, {Ott}, {Radice}, {Roberts},
  {Schnetter}, \& {Haas}}]{2015Natur.528..376M}
{M{\"o}sta}, P., {Ott}, C.~D., {Radice}, D., {et~al.} 2015, \nat, 528, 376

\bibitem[{{M{\"o}sta} {et~al.}(2014){M{\"o}sta}, {Richers}, {Ott}, {Haas},
  {Piro}, {Boydstun}, {Abdikamalov}, {Reisswig}, \&
  {Schnetter}}]{2014ApJ...785L..29M}
{M{\"o}sta}, P., {Richers}, S., {Ott}, C.~D., {et~al.} 2014, \apjl, 785, L29

\bibitem[{{Nishimura} {et~al.}(2017){Nishimura}, {Hirschi}, {Rauscher},
  {Murphy}, \& {Cescutti}}]{2017arXiv170100489N}
{Nishimura}, N., {Hirschi}, R., {Rauscher}, T., {Murphy}, A.~S.~J., \&
  {Cescutti}, G. 2017, ArXiv e-prints, arXiv:1701.00489

\bibitem[{{Nishimura} {et~al.}(2012){Nishimura}, {Kajino}, {Mathews},
  {Nishimura}, \& {Suzuki}}]{2012PhRvC..85d8801N}
{Nishimura}, N., {Kajino}, T., {Mathews}, G.~J., {Nishimura}, S., \& {Suzuki},
  T. 2012, \prc, 85, 048801

\bibitem[{{Nishimura} {et~al.}(2016){Nishimura}, {Podoly{\'a}k}, {Fang}, \&
  {Suzuki}}]{2016PhLB..756..273N}
{Nishimura}, N., {Podoly{\'a}k}, Z., {Fang}, D.-L., \& {Suzuki}, T. 2016,
  Physics Letters B, 756, 273

\bibitem[{{Nishimura} {et~al.}(2015){Nishimura}, {Takiwaki}, \&
  {Thielemann}}]{2015ApJ...810..109N}
{Nishimura}, N., {Takiwaki}, T., \& {Thielemann}, F.-K. 2015, \apj, 810, 109

\bibitem[{{Nomoto} {et~al.}(2013){Nomoto}, {Kobayashi}, \&
  {Tominaga}}]{2013ARA&A..51..457N}
{Nomoto}, K., {Kobayashi}, C., \& {Tominaga}, N. 2013, \araa, 51, 457

\bibitem[{{Obergaulinger} {et~al.}(2009){Obergaulinger}, {Cerd{\'a}-Dur{\'a}n},
  {M{\"u}ller}, \& {Aloy}}]{2009A&A...498..241O}
{Obergaulinger}, M., {Cerd{\'a}-Dur{\'a}n}, P., {M{\"u}ller}, E., \& {Aloy},
  M.~A. 2009, \aap, 498, 241

\bibitem[{{Papish} \& {Soker}(2012)}]{2012MNRAS.421.2763P}
{Papish}, O., \& {Soker}, N. 2012, \mnras, 421, 2763

\bibitem[{{Rauscher} \& {Thielemann}(2000)}]{2000ADNDT..75....1R}
{Rauscher}, T., \& {Thielemann}, F.-K. 2000, Atomic Data and Nuclear Data
  Tables, 75, 1

\bibitem[{{Rembiasz} {et~al.}(2016){Rembiasz}, {Obergaulinger},
  {Cerd{\'a}-Dur{\'a}n}, {M{\"u}ller}, \& {Aloy}}]{2016MNRAS.456.3782R}
{Rembiasz}, T., {Obergaulinger}, M., {Cerd{\'a}-Dur{\'a}n}, P., {M{\"u}ller},
  E., \& {Aloy}, M.~A. 2016, \mnras, 456, 3782

\bibitem[{{Roederer} {et~al.}(2016){Roederer}, {Mateo}, {Bailey}, {Song},
  {Bell}, {Crane}, {Loebman}, {Nidever}, {Olszewski}, {Shectman}, {Thompson},
  {Valluri}, \& {Walker}}]{2016AJ....151...82R}
{Roederer}, I.~U., {Mateo}, M., {Bailey}, III, J.~I., {et~al.} 2016, \aj, 151,
  82

\bibitem[{{Sawai} \& {Yamada}(2014)}]{2014ApJ...784L..10S}
{Sawai}, H., \& {Yamada}, S. 2014, \apjl, 784, L10

\bibitem[{{Sawai} \& {Yamada}(2016)}]{2016ApJ...817..153S}
---. 2016, \apj, 817, 153

\bibitem[{{Sawai} {et~al.}(2013){Sawai}, {Yamada}, \&
  {Suzuki}}]{2013ApJ...770L..19S}
{Sawai}, H., {Yamada}, S., \& {Suzuki}, H. 2013, \apjl, 770, L19

\bibitem[{{Shibata} {et~al.}(2006){Shibata}, {Liu}, {Shapiro}, \&
  {Stephens}}]{2006PhRvD..74j4026S}
{Shibata}, M., {Liu}, Y.~T., {Shapiro}, S.~L., \& {Stephens}, B.~C. 2006, \prd,
  74, 104026

\bibitem[{{Sneden} {et~al.}(1996){Sneden}, {McWilliam}, {Preston}, {Cowan},
  {Burris}, \& {Armosky}}]{1996ApJ...467..819S}
{Sneden}, C., {McWilliam}, A., {Preston}, G.~W., {et~al.} 1996, \apj, 467, 819

\bibitem[{{Takiwaki} {et~al.}(2009){Takiwaki}, {Kotake}, \&
  {Sato}}]{2009ApJ...691.1360T}
{Takiwaki}, T., {Kotake}, K., \& {Sato}, K. 2009, \apj, 691, 1360

\bibitem[{{Takiwaki} {et~al.}(2016){Takiwaki}, {Kotake}, \&
  {Suwa}}]{2016MNRAS.461L.112T}
{Takiwaki}, T., {Kotake}, K., \& {Suwa}, Y. 2016, \mnras, 461, L112

\bibitem[{{Tsujimoto} \& {Nishimura}(2015)}]{2015ApJ...811L..10T}
{Tsujimoto}, T., \& {Nishimura}, N. 2015, \apjl, 811, L10

\bibitem[{{Wallner} {et~al.}(2015){Wallner}, {Faestermann}, {Feige},
  {Feldstein}, {Knie}, {Korschinek}, {Kutschera}, {Ofan}, {Paul}, {Quinto},
  {Rugel}, \& {Steier}}]{2015NatCo...6E5956W}
{Wallner}, A., {Faestermann}, T., {Feige}, J., {et~al.} 2015, Nature
  Communications, 6, 5956

\bibitem[{{Wanajo} {et~al.}(2014){Wanajo}, {Sekiguchi}, {Nishimura}, {Kiuchi},
  {Kyutoku}, \& {Shibata}}]{2014ApJ...789L..39W}
{Wanajo}, S., {Sekiguchi}, Y., {Nishimura}, N., {et~al.} 2014, \apjl, 789, L39

\bibitem[{{Wehmeyer} {et~al.}(2015){Wehmeyer}, {Pignatari}, \&
  {Thielemann}}]{2015MNRAS.452.1970W}
{Wehmeyer}, B., {Pignatari}, M., \& {Thielemann}, F.-K. 2015, \mnras, 452, 1970

\bibitem[{{Winteler} {et~al.}(2012){Winteler}, {K{\"a}ppeli}, {Perego},
  {Arcones}, {Vasset}, {Nishimura}, {Liebend{\"o}rfer}, \&
  {Thielemann}}]{2012ApJ...750L..22W}
{Winteler}, C., {K{\"a}ppeli}, R., {Perego}, A., {et~al.} 2012, \apjl, 750, L22

\bibitem[{{Woosley} \& {Weaver}(1995)}]{1995ApJS..101..181W}
{Woosley}, S.~E., \& {Weaver}, T.~A. 1995, \apjs, 101, 181

\bibitem[{{Wu} {et~al.}(2016){Wu}, {Fern{\'a}ndez}, {Mart{\'{\i}}nez-Pinedo},
  \& {Metzger}}]{2016MNRAS.463.2323W}
{Wu}, M.-R., {Fern{\'a}ndez}, R., {Mart{\'{\i}}nez-Pinedo}, G., \& {Metzger},
  B.~D. 2016, \mnras, 463, 2323

\bibitem[{{Yamada} \& {Sawai}(2004)}]{2004ApJ...608..907Y}
{Yamada}, S., \& {Sawai}, H. 2004, \apj, 608, 907

\end{thebibliography}
\end{document}